\documentclass[twocolumn,aps,floatfix]{revtex4}
\usepackage{amssymb}
\usepackage{graphicx}

\begin{document}
%\draft
\title{On the stability of Bose-Fermi mixtures}

\author{T. Karpiuk,$\,^1$ M. Brewczyk,$\,^1$
        M. Gajda,$\,^2$ and K. Rz{\c a}\.zewski$\,^3$}

\affiliation{\mbox{$^1$ Instytut Fizyki Teoretycznej, Uniwersytet w Bia{\l}ymstoku,
                        ulica Lipowa 41, 15-424 Bia{\l}ystok, Poland}  \\
\mbox{$^2$ Instytut Fizyki PAN, Aleja Lotnik\'ow 32/46, 02-668 Warsaw, Poland} \\
\mbox{$^3$ Centrum Fizyki Teoretycznej PAN, Aleja Lotnik\'ow 32/46, 02-668 Warsaw,
           Poland}  }

\date{\today}

\begin{abstract}
We consider the stability of a mixture of degenerate Bose and Fermi
gases. Even though the bosons effectively repel each other the mixture
can still collapse provided the Bose and Fermi gases attract each other
strongly enough. For a given number of atoms and the strengths of the
interactions between them we find the geometry of a maximally compact 
trap that supports the stable mixture. We compare a simple analytical
estimation for the critical axial frequency of the trap with results 
based on the numerical solution of hydrodynamic equations for Bose-Fermi 
mixture.

\end{abstract}

\maketitle

Experimental realizations of mixtures of degenerate atomic gases have
opened exceptional possibilities to investigate fundamental many-body
quantum phenomena. The reason is that in such systems there exists a powerful 
tool (Feshbach resonances) to control the strength of the interaction between 
atoms. Magnetically tuned scattering resonances allow for changing the magnitude 
as well as the sign of the scattering length which at low temperatures fully 
determines the atomic interactions. Recently, mixtures of fermionic gases 
turned out to be a successful way through in obtaining a molecular Bose-Einstein 
condensates (BEC) in thermal equilibrium \cite{molBEC}. Such molecular condensates are
good starting points to experimental study of BEC-BCS (Bardeen-Cooper-Schrieffer)
crossover region; subject under intensive theoretical investigation. Two-component 
Fermi gases enabled also getting into a regime of strongly interacting fermionic 
systems \cite{Thomas}, here confirmed by the observation of anisotropic expansion 
of the gas when released from a trap. In Ref. \cite{Grimm} the pairing gap in a 
strongly interacting two-component gas of fermionic $^6$Li atoms has been measured 
showing that the system was already brought in a superfluid state.

Another idea of modifying the interactions between fermions is to immerse Fermi 
atoms into a Bose gas. In this way a degenerate gas of fermionic potassium 
($^{40}$Ka) was forced to collapse by attractive interaction with rubidium 
($^{87}$Rb) Bose-Einstein condensate \cite{collapse}. As it was shown in this
experiment, a sufficiently large number of atoms is required to bring a Bose-Fermi
mixture to collapse. This result suggests that increasing the attraction between 
the bosonic and fermionic atoms one can induce the effective attractive interaction 
between fermions. Increasing the effective fermion-fermion attraction could, on the 
other hand, lead to a superfluidity in fermionic component. This kind of superfluidity 
resembles what happens in superconductors where phonon-induced interaction between 
electrons becomes attractive.

In Ref. \cite{BFsol} it has been shown (although in a one-dimensional case)
that for strong enough attraction between bosons and fermions a Bose-Fermi 
mixture enters a new phase where both Bose and Fermi components become effectively 
attractive gases. One of the characteristics of this regime is that it becomes 
possible to generate bright solitons which are two-component single-peak structures 
with larger number of bosons than fermions. Since in a three-dimensional space such 
a mixture might show an instability that could destroy it, it is necessary to 
investigate in detail the stability-instability crossover. The structure and 
the instability of the boson-fermion mixtures have been theoretically studied in 
Ref. \cite{Roth}, although the analysis was restricted to spherically symmetric systems 
with equal numbers of atoms in both components. The authors determine, by fitting 
the numerical data, the expressions for critical numbers of atoms as a function of 
scattering lengths. In this Letter we investigate the border of the stability region 
for any trap, find analytic formula for the critical radial trap frequency and 
compare it with results of numerical solution of hydrodynamic equations describing 
the Bose-Fermi mixture.

In the mean-field approximation the Bose-Fermi mixture of $N_B$ bosons
and $N_F$ fermions can be described in terms of atomic orbitals 
\cite{FF,BFsol} that fulfill the following set of equations 
($j = 1,2,...,N_F$)
\begin{eqnarray}
&&i\hbar\,\frac{\partial\varphi^{(B)}}{\partial t} = -\frac{\hbar^2}{2 m_B}
\nabla^2 \varphi^{(B)} + V_{trap}^{(B)} \, \varphi^{(B)}  \nonumber  \\
&&+\; g_B\, N_B\, |\varphi^{(B)}|^2 \, \varphi^{(B)}
+ g_{BF} \sum_{i=1}^{N_F} |\varphi_i^{(F)}|^2 \, \varphi^{(B)}
\nonumber  \\
&&i\hbar\,\frac{\partial\varphi_j^{(F)}}{\partial t} = -\frac{\hbar^2}{2 m_F}
\nabla^2 \varphi_j^{(F)} + V_{trap}^{(F)} \, \varphi_j^{(F)}
\nonumber  \\
&&+\; g_{BF}\, N_B\, |\varphi^{(B)}|^2 \, \varphi_j^{(F)}   
\label{orbitals}
\end{eqnarray}
and determine the many-body wave function of the mixture, which is of
the form
\begin{eqnarray}
&&\Psi ({\bf x}_1,...,{\bf x}_{N_B};{\bf y}_1,...,{\bf y}_{N_F}) =
\prod_{i=1}^{N_B} \varphi^{(B)}({\bf x}_i)
\nonumber \\
&&\times \frac{1}{\sqrt{N_F!}} \left |
\begin{array}{lllll}
\varphi_1^{(F)}({\bf y}_1) & . & . & . & \varphi_1^{(F)}({\bf y}_{N_F}) \\
\phantom{aa}. &  &  &  & \phantom{aa}. \\
\phantom{aa}. &  &  &  & \phantom{aa}. \\
\phantom{aa}. &  &  &  & \phantom{aa}. \\
\varphi_{N_F}^{(F)}({\bf y}_1) & . & . & . & \varphi_{N_F}^{(F)}({\bf y}_{N_F})
\end{array}
\right |    %\times   \nonumber \\ \nonumber \\  \;.
\label{wavefunction}
\end{eqnarray}
We assume the contact interaction between atoms with $g_B$ and $g_{BF}$ being 
the coupling constants for the boson-boson and boson-fermion interactions, 
respectively.

\begin{figure}[thb]
\resizebox{3.1in}{2.1in} {\includegraphics{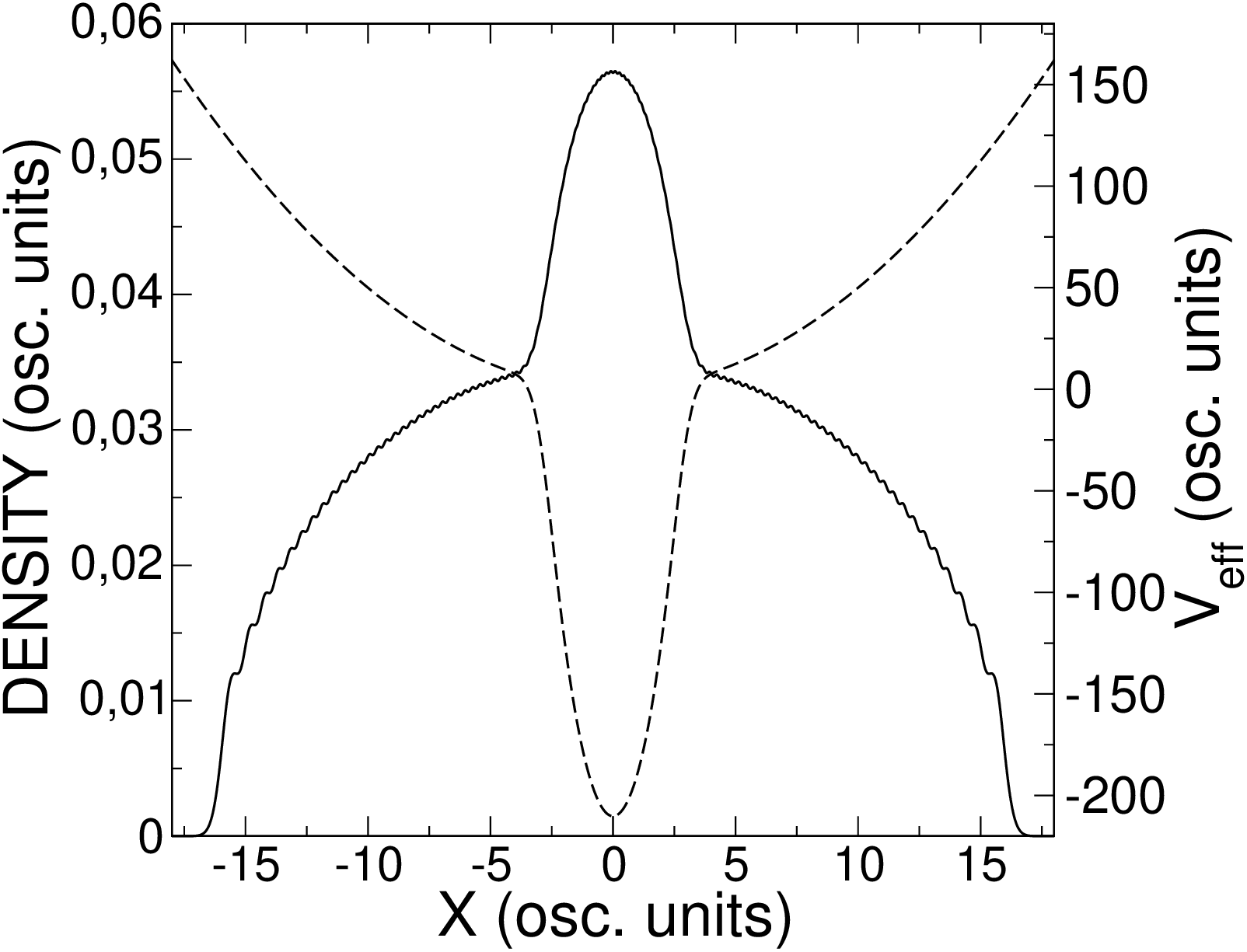}}
\caption{Fermionic density (solid line) normalized to one for a one-dimensional
Bose-Fermi mixture of $1000$ bosons and $100$ fermions. The strength of effective 
repulsion for bosons is $g_B=0.0163$ and the attraction between bosons and fermions
is determined by $g_{BF}= -0.5$ in oscillatory units. The dashed line shows the
effective potential for fermions.} 
\label{potden}
\end{figure}

Eqs. (\ref{orbitals}) imply that fermions move under the influence of the 
effective potential which is the sum of the harmonic trap and the potential 
that originates in the presence of bosons
\begin{eqnarray}
V_{eff} = V_{trap}^{(F)} + g_{BF} N_B |\,\varphi^{(B)}|^2   \;.
\label{effective}
\end{eqnarray}
In Fig. \ref{potden} we plot this potential (dashed line) as well as the
fermionic density (the sum of orbital densities) in the case of one-dimensional
system. It is clear from (\ref{effective}) that when the bosons and fermions
attract each other strongly enough, the second term in (\ref{effective})
generates the well at the bottom of the harmonic trap (see Fig. \ref{potden}).
Hence, one can distinguish  between fermions captured by the bosons (those
with energies below zero) and others with higher energies that built broad
fermionic basis.

To estimate the number of fermions pulled inside the bosonic cloud we use 
the Thomas-Fermi approximation. Since we consider the $^{87}$Rb and $^{40}$K
atoms in a magnetic trap in their doubly spin polarized states, we utilize 
the relation $m_F (\omega_\alpha^F)^2 = m_B (\omega_\alpha^B)^2$ (with $\alpha$ 
equal to $x, y$, and $z$) and rewrite the potential (\ref{effective}) in the 
following form
\begin{eqnarray}
V_{eff} = \left\{  \begin{array}{lll}
-\frac{|g_{BF}|}{g_B} \mu_B + (\frac{|g_{BF}|}{g_B}+1)\, V_{trap}^{(B)}  
 &  & \mbox{inside bosons}  \\
 &  & \\
V_{trap}^{(F)} & & \mbox{\hfill otherwise}
\end{array}
\right.
\label{potential}
\end{eqnarray}
where $\mu_B$ is the chemical potential for the Bose-Einstein condensate
in the Thomas-Fermi limit \cite{Stringari}
\begin{eqnarray}
\mu_B = \frac{\hbar \omega_{ho}}{2} \left(\frac{15\, a_B N_B}{a_{ho}}
\right)^{2/5}   \;.
\label{chempot}
\end{eqnarray}
Here, $a_{ho}=\sqrt{\hbar/(m_B \omega_{ho})}$ is the harmonic oscillator 
length and $\omega_{ho}=(\omega_x \omega_y \omega_z)^{1/3}$ is the geometric
average of the oscillator frequencies. The scattering length $a_B$ and
the interaction strength $g_B$ are related through
$g_B=4\pi\hbar^2a_B/m_B$.

Now, the number of fermions captured by bosons is just the number of states 
existing in the potential well defined by the upper branch of (\ref{potential}).
Since this potential acts inside the bosonic cloud, one has to count only states  
with energies below $\frac{|g_{BF}|}{g_B} \mu_B$ with respect to the minimum 
of the well. The well itself is the harmonic potential with frequencies
multiplied by factors $\sqrt{|g_{BF}|/g_B+1}$ in comparison with original
ones for bosons. The total number of states in a harmonic potential with 
the energy less than $\epsilon$ is given by 
\cite{Pethick}
\begin{eqnarray}
G(\epsilon) = \frac{\epsilon^3}{6 \hbar^3\, \omega_x \omega_y \omega_z}  \;.
\label{states}
\end{eqnarray}
Therefore, the number of fermions pulled inside the bosons equals
\begin{eqnarray}
N_F^{in} = \frac{\mu^3_B}{6 \hbar^3\, \omega_x \omega_y \omega_z}
\; \frac{(|g_{BF}|/g_B)^3}{(|g_{BF}|/g_B+1)^{3/2}}
\label{NF}
\end{eqnarray}
and after inserting the chemical potential given by (\ref{chempot}) this
number becomes
\begin{eqnarray}
N_F^{in} = \frac{1}{48} \left(\frac{15\, a_B N_B}{a_{ho}}\right)^{6/5}
\frac{(|g_{BF}|/g_B)^3}{(|g_{BF}|/g_B+1)^{3/2}}   \;.
\label{NF1}
\end{eqnarray}

Having calculated the number of fermions captured by bosons one can separate
off the bosonic component. The energy of the Bose component of a Bose-Fermi 
mixture in the mean-field approximation reads
\begin{eqnarray}
E &=& \int  \left\{ \frac{\hbar^2}{2 m_B} |\nabla \Psi_B|^2 +
\frac{m_B}{2} [ \omega_{\bot}^2 (x^2+y^2) + \omega_z^2 z^2 ]\, |\Psi_B|^2  
\right.   \nonumber  \\
&+& \left. \frac{1}{2}\, g_B |\Psi_B|^4 + g_{BF} |\Psi_B|^2 n_F^{in}  \right\} 
\, d^{\,3} r    \;,
\label{energy}
\end{eqnarray}
where $\Psi_B=N_B \varphi^{(B)}$ and $n_F^{in}({\bf r})$ is the fermionic density 
drown into the bosonic cloud. Simplifying the description of the system we introduce 
the Gaussian variational ansatz for the condensate wave function as well as for the 
fermionic density
\begin{eqnarray}
\Psi_B(\rho,z) &=& \frac{\sqrt{N_B}}{\pi^{3/4} w \sqrt{v}} 
\exp{\left[-\frac{1}{2}\left(\frac{\rho^2}{w^2}+\frac{z^2}{v^2}\right)\right]}
\nonumber  \\
\sqrt{n_F^{in}(\rho,z)} &=& \frac{\sqrt{N_F^{in}}}{\pi^{3/4} w \sqrt{v}} 
\exp{\left[-\frac{1}{2}\left(\frac{\rho^2}{w^2}+\frac{z^2}{v^2}\right)\right]}
,
\label{ansatz}
\end{eqnarray}
where the radial and the axial widths, $w$ and $v$ respectively, are
assumed to be the same for both species. This assumption allows for elimination 
of the fermionic degrees of freedom. Now, the energy of bosons is given by the 
expression (in oscillatory units based on the axial frequency)
\begin{eqnarray}
\frac{E}{N_B} &=& \frac{1}{2} \left(\frac{1}{2v^2}+\frac{1}{w^2}
\right) + \frac{1}{4} \left(v^2 + 2 \beta^2 w^2\right)   \nonumber   \\
&+& \frac{A}{\sqrt{2\pi \beta}}  \frac{1}{v w^2}   \;,
\label{energy1}
\end{eqnarray}
where
\begin{eqnarray}
A = \frac{1}{a_\bot}  \left (a_B N_B + a_{BF} N_F^{in} \frac{m_B}{\mu} \right) \;,
\label{A}
\end{eqnarray}
the aspect ratio $\beta=\omega_{\bot}/\omega_z$, and the reduced mass
$\mu=m_B m_F /(m_B+m_F)$. The first term in the formula (\ref{energy1}) 
is the kinetic energy of the Bose cloud, the second one describes the 
harmonic trapping potential whereas the last term covers the interaction 
between atoms. Since the bosons and fermions attract each other, $a_{BF}$ 
is negative and when this attraction is strong enough, i.e. in the region 
of parameters close to the stability border the constant $A$ becomes negative. 
Therefore, on the edge of the mixture stability the expression (\ref{energy1}) 
can be treated just as the energy of a Bose gas of effectively attractive atoms.

We now find, for a given trap, the maximum strength of attraction between
bosons which still allows for the existence of the stable Bose-Einstein
condensate. We look for the minimum of the energy (\ref{energy1}). The
necessary condition for that turns to the set of following equations
for the widths
\begin{eqnarray}
&& -1 - \frac{2 A}{\sqrt{2\pi \beta}} \frac{1}{v} + \beta^2 w^4 =0   \\
&& -\frac{1}{2} + \frac{v^4}{2} - \frac{A}{\sqrt{2\pi \beta}} \frac{v}{w^2}= 0 \;.
\label{first}
\end{eqnarray}
Calculating $w^2$ from the first equation and inserting it in the
second one leads to the equation
\begin{eqnarray}
\frac{1}{2} (-1 + v^4) =  \frac{A \sqrt{\beta /2\pi}\, v}
{\sqrt{1+\frac{2 A}{\sqrt{2\pi \beta}} \frac{1}{v}}}   \;.
\label{second}
\end{eqnarray}
Since $A$ is negative, the Eq. (\ref{second}) has a solution
when the maximum of the right hand side of (\ref{second}) as a
function of the width $v$ is bigger than the value of the left
hand side of (\ref{second}) at the point of this  maximum. Then
the critical value of the strength $A$ fulfills the biquadratic 
equation
\begin{eqnarray}
\frac{81}{8 \pi^2 \beta^2} A^4 + \frac{3\sqrt{3}}{2\pi} A^2 -\frac{1}{2} = 0
\label{Aequ}
\end{eqnarray}
and the solution
\begin{eqnarray}
|A_{cr}| = \frac{\sqrt{2\pi}}{3^{5/4}}\,
\sqrt{-\beta^2 + \sqrt{3 \beta^2 + \beta^4}} \equiv f(\beta)
\label{fbeta}
\end{eqnarray}
generalizes well known result for a spherically symmetric trap 
\cite{Stringari}.

Equating (\ref{A}) and (\ref{fbeta}) one obtains the condition for
the parameters of the maximally compact trap which still holds the 
Bose-Fermi mixture of a given number of atoms and the interaction
strengths
\begin{eqnarray}
|a_{BF}|\, \widetilde{N_F^{in}}\, \frac{m_B}{\mu}\,  \beta^{2/5}\, 
\omega_z^{3/5} - a_B N_B   =   \sqrt{\frac{\hbar}{m_B}}\,
\frac{f(\beta)}{\sqrt{\beta \omega_z}}    \;,
\label{condition}
\end{eqnarray}
where
\begin{eqnarray}
\widetilde{N_F^{in}} = \frac{1}{48} \left( \sqrt{\frac{m_B}{\hbar}} 15\, a_B N_B
\right)^{6/5} \!\! \frac{(|g_{BF}|/g_B)^3}{(|g_{BF}|/g_B+1)^{3/2}}  \;.
\label{NF2}
\end{eqnarray}
Solving the condition (\ref{condition}) determines the critical trap 
frequencies. It turns out that both terms on the left hand side of 
(\ref{condition}) are much bigger than the term on the right hand side. 
Therefore, the critical axial trap frequency is given by
\begin{eqnarray}
\omega_z^{cr} = \left(
\frac{a_B N_B}{|a_{BF}|\, \widetilde{N_F^{in}}}\,  \frac{\mu}{m_B} \right)^{5/3}  
\!\! \frac{1}{\beta^{2/3}}   \;.
\label{critical}
\end{eqnarray}

The formula (\ref{critical}) can be also regarded as a way of determining
the mutual scattering length $a_{BF}$ when one knows the critical number of
bosons (and at the same time the number of drown in fermions is given by
(\ref{NF1})) for a particular set of trap parameters. In such a case one has 
to solve the following equation
\begin{eqnarray}
C\, \omega_z^{3/5} \beta^{2/5}\, |a_{BF}|^4 = (D\, |a_{BF}| + 1)^{3/2}  \;,
\label{aBF}
\end{eqnarray}
where
\begin{eqnarray}
&&C = \frac{1}{384 N_B^{cr}} \frac{1}{a_B^4} \left(\frac{m_B}{\mu}\right)^4
\left( \sqrt{\frac{m_B}{\hbar}} 15\, a_B N_B^{cr} \right)^{6/5}  
\nonumber  \\
&&D = \frac{1}{2 a_B}  \frac{m_B}{\mu}    \;.
\label{CD}
\end{eqnarray}

\begin{figure}[bht]
\resizebox{3.1in}{2.3in} {\includegraphics{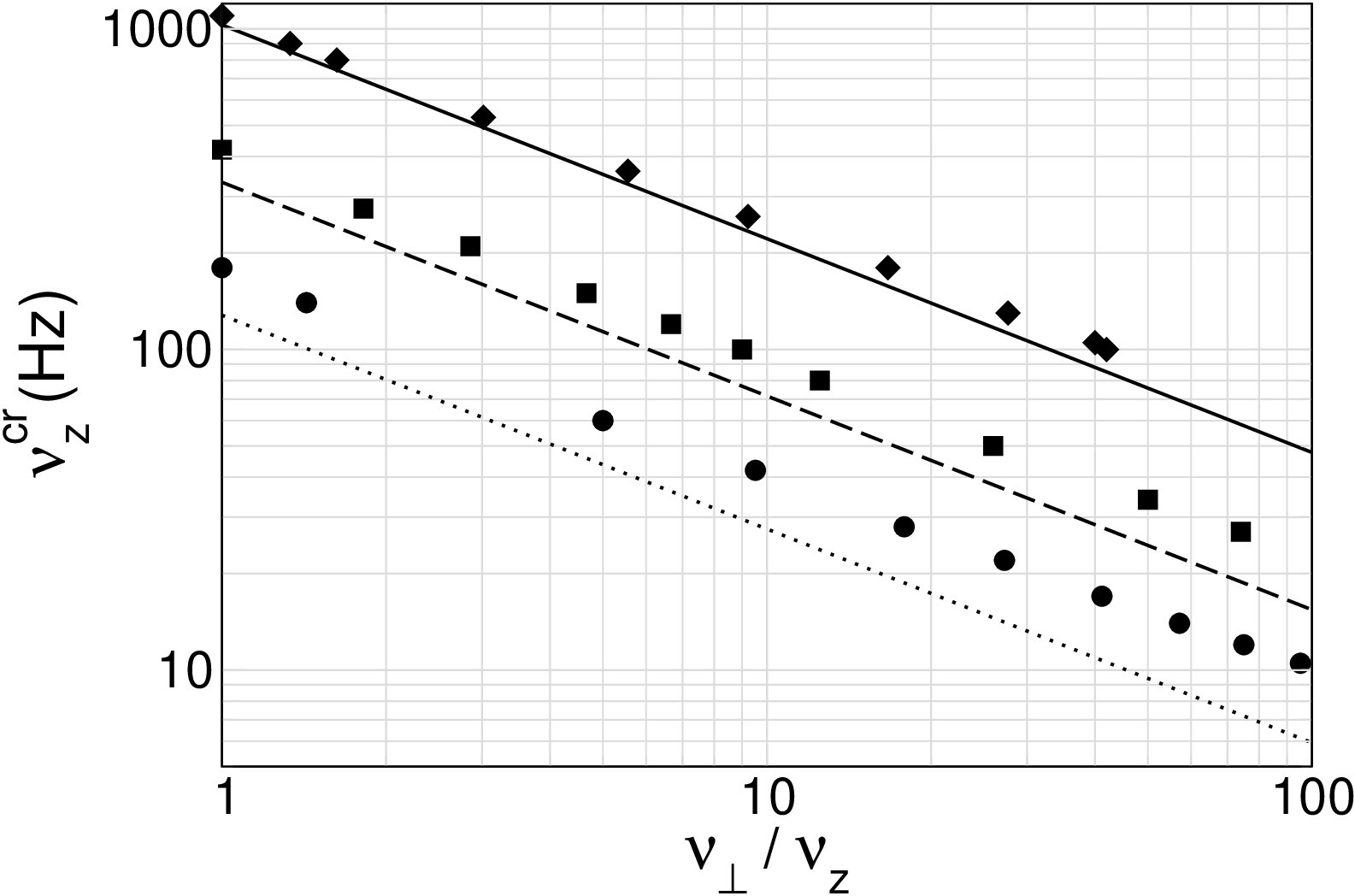}}
\caption{Critical axial frequency as a function of the aspect ratio
for the mixture of $10^4$ bosons and $10^4$ fermions. Points and lines 
come from the numerical solution of a set of hydrodynamic equations 
and from the analytical formula (\ref{critical}), respectively. Solid 
circles (dotted line), squares (dashed line), and diamonds (solid line) 
correspond to the mutual scattering length 
$a_{BF}=-21.7$\,nm, $-17.6$\,nm, and $-13.8$\,nm, successively.} 
\label{10k}
\end{figure}

\begin{figure}[thb]
\resizebox{3.1in}{2.3in} {\includegraphics{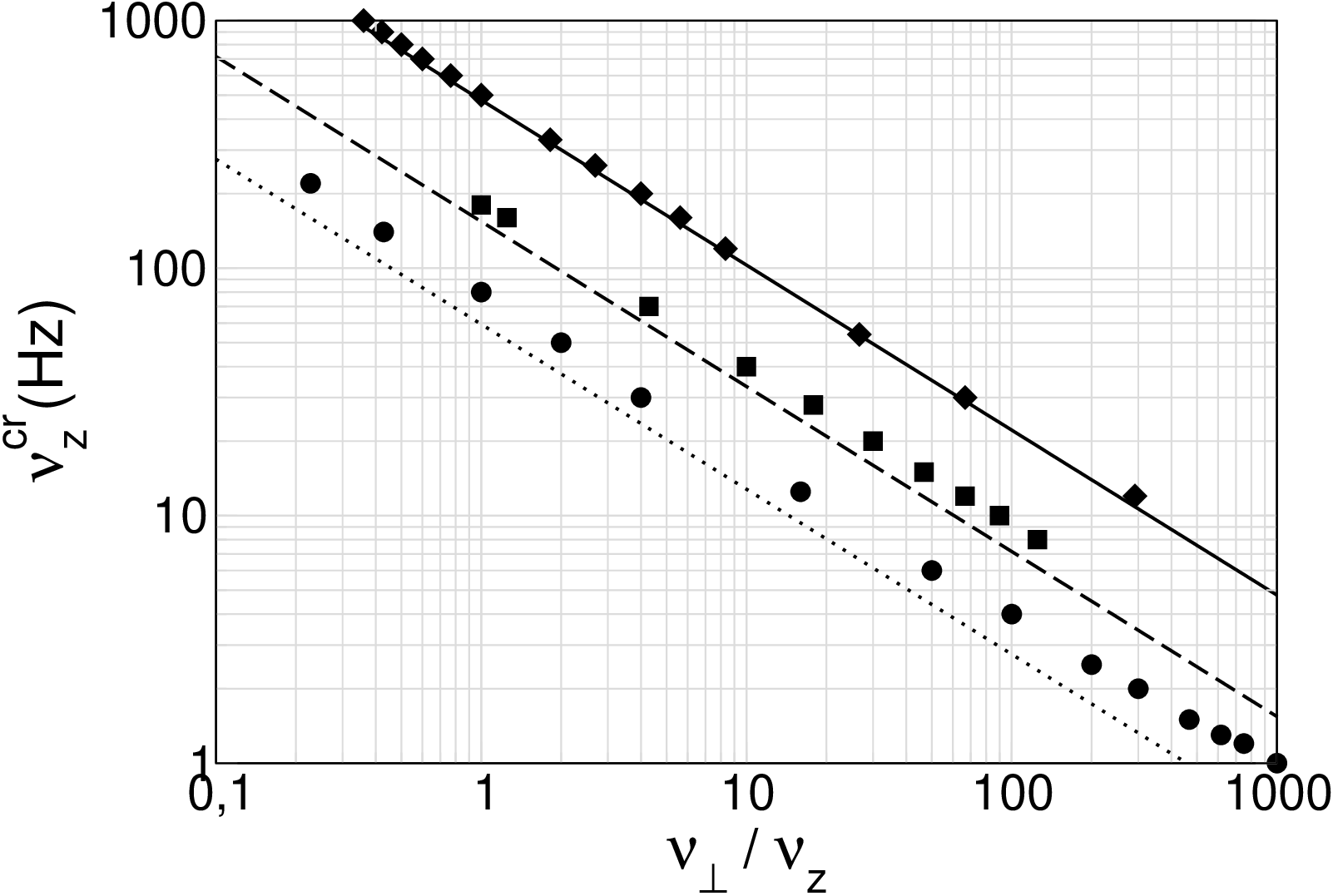}}
\caption{The same as in Fig. \ref{10k} but for larger number of atoms,
here $N_B=N_F=10^5$.} 
\label{100k}
\end{figure}

In Figs. \ref{10k} and \ref{100k} we plot $\nu_z^{cr}=\omega_z^{cr}/(2\pi)$ 
as a function of the aspect ratio, calculated based on the formula 
(\ref{critical}). Since there is a controversy over the value of the scattering 
length $a_{BF}$ \cite{collapse,Simoni,Jin}, we included three different values 
of it. We show also points obtained from numerical integration of a set of 
equations that are a hydrodynamic version of Eqs. (\ref{orbitals}). The hydrodynamic
equations can be derived from a set of equations for reduced density matrices 
(see Ref. \cite{Tomek}) after making a local equilibrium assumption for fermions 
(i.e., utilizing the Thomas-Fermi approximation) and calculating the interaction 
between bosons and fermions within the mean-field approach. The basic ''objects'' 
in the hydrodynamic approximation are then the condensate wave function and the 
fermionic density and velocity fields. Hence, the equation for the condensate 
wave function looks like that in a set of Eqs. (\ref{orbitals}) but with the 
last term equal to $g_{BF} n_{F} \varphi^{(B)}$ (with $n_{F}$ being the fermionic
density). On the other hand, the equations describing fermions are the usual 
hydrodynamic equations, i.e., the continuity equation and the Euler-type equation 
of motion.

Figs. \ref{10k} and \ref{100k} show good agreement between the formula 
(\ref{critical}) and the numerical points for weaker boson-fermion attraction
(i.e., less negative scattering length $a_{BF}$). The growing discrepancy
for stronger attraction means that the number of fermions pulled inside
bosons is overestimated. Indeed, the formula (\ref{NF1}) derived based on
the Thomas-Fermi approximation for bosons, does not account for the effect
of shrinking the bosonic cloud while capturing more and more fermions.

In conclusion, we have analyzed the Bose-Fermi mixture in a regime
of parameters, where both gases start to behave as a systems of
effectively attractive atoms and the existence of the mixture is in 
danger due to the possible collapse. We found an analytical formula 
which determines the parameters of maximally compact trap allowing for 
the stable mixture. We compare this estimate with the results obtained 
by solving numerically the hydrodynamic equations describing the 
Bose-Fermi mixture. It turns out that the collapse is very sensitive to 
the value of the scattering length $a_{BF}$ and therefore could help
to determine this scattering length.

\acknowledgments 
We thank S. Ospelkaus-Schwarzer and K. Bongs for helpful discussions.
The authors acknowledge support by the Polish Ministry of Scientific 
Research Grant Quantum Information and Quantum Engineering 
No. PBZ-MIN-008/P03/2003.

\end{document}